\documentclass[conference]{IEEEtran}
\IEEEoverridecommandlockouts

\usepackage{cite}
\ifCLASSINFOpdf
  \usepackage[pdftex]{graphicx}
  \graphicspath{{../pdf/}{../jpeg/}}
  \DeclareGraphicsExtensions{.pdf,.jpeg,.png}
\else
  \usepackage[dvips]{graphicx}
  \graphicspath{{../eps/}}
  \DeclareGraphicsExtensions{.eps}
\fi

\usepackage{amsmath,amssymb}
\usepackage{mathtools}
\usepackage{bm}
\usepackage[caption=false,font=footnotesize]{subfig}

\usepackage{siunitx}

\DeclareMathOperator{\rect}{rect}

\DeclareMathOperator{\qfunc}{Q}
\DeclareMathOperator{\Pe}{Pe}
\providecommand{\Deff}{D_\textrm{eff}}

\providecommand{\veff}{v_\textrm{eff}}

\providecommand{\vmean}{\veff}

\providecommand{\vmax}{v_\textrm{max}}

\providecommand{\dV}{\mathrm{d}V}

\providecommand{\tmax}{t_\mathrm{max}}
\providecommand{\pmax}{p_\mathrm{max}}

\providecommand{\Pob}{P_\mathrm{ob}}
\providecommand{\Nob}{N_\mathrm{ob}}
\providecommand{\Pobd}{P_{\mathrm{ob},\mathrm{d}}}
\providecommand{\Pobf}{P_{\mathrm{ob},\mathrm{f}}}
\providecommand{\Pobfp}{P^\bullet_{\mathrm{ob},\mathrm{f}}}
\providecommand{\Ntx}{N_\textsc{tx}}
\providecommand{\RX}{\textsc{rx}}

\providecommand{\Vrx}{V_\RX}
\providecommand{\pgrad}{\partial_x P}
\providecommand{\Nn}{N_\mathrm{n}}
\providecommand{\figspacelower}{\vspace*{-4mm}}
\providecommand{\figspaceupper}{\vspace*{-2mm}}
\providecommand{\Arx}{A_\RX}
\providecommand{\ahat}{\hat{a}}
\providecommand{\Nobmean}{\overline{\Nob}}

\begin{document}

\title{Modeling Duct Flow for Molecular Communication\vspace*{-4mm}}

\author{\IEEEauthorblockN{Wayan Wicke\IEEEauthorrefmark{1},
        Tobias Schwering\IEEEauthorrefmark{1},
        Arman Ahmadzadeh\IEEEauthorrefmark{1},
        Vahid Jamali\IEEEauthorrefmark{1},
        Adam Noel\IEEEauthorrefmark{2},
        and Robert Schober\IEEEauthorrefmark{1}
\IEEEauthorblockA{\IEEEauthorrefmark{1}Institute for Digital Communications, University of Erlangen-Nuremberg
}
\IEEEauthorblockA{\IEEEauthorrefmark{2}School of Engineering, University of Warwick, Coventry, UK\vspace*{-6mm}}%
}}


\maketitle

\begin{abstract}
Active transport such as fluid flow is sought in molecular communication to extend coverage, improve reliability, and mitigate interference.
Flow models are often over-simplified, assuming one-dimensional diffusion with constant drift.
However, diffusion and flow are usually encountered in three-dimensional bounded environments where the flow is highly non-uniform such as in blood vessels or microfluidic channels.
For a qualitative understanding of the relevant physical effects inherent to these channels, based on the P\'eclet number and the transmitter-receiver distance, we study when simplified models of uniform flow and advection-only transport are applicable.
For these two regimes, analytical expressions for the channel impulse response are derived and validated by particle-based simulation.
Furthermore, as advection-only transport is typically overlooked and hence not analyzed in the molecular communication literature, we evaluate the symbol error rate for exemplary on-off keying as performance metric.
\end{abstract}


\section{Introduction}
Using molecules for conveying digital messages has recently been recognized as a key communication strategy for nanoscale devices such as artificial cells cooperatively fighting a disease \cite{farsad_comprehensive_2016}.
As the entities involved in this molecular communication are in the nano- and microscale, diffusion plays a significant role in the propagation of messages \cite{farsad_comprehensive_2016}.

However, diffusion has a limited effective range that renders molecular communication inefficient over extended distances.
This limitation can be overcome by exploiting fluid flow in addition to diffusion.
For example, in blood vessels it is the interplay of fluid flow and diffusion that governs the supply of oxygen from the lungs to tissues.
Consequently, the molecular communication literature has considered basic models of these fundamental mechanisms \cite{farsad_comprehensive_2016}.
In particular, the basic channel characteristics of diffusion in three-dimensional (3D) unbounded space with uniform flow in the context of molecular communication have been investigated for example in \cite{noel_improving_2014}.
Such a model might be applicable when the boundaries are far from the nanonetwork.
Our previous work \cite{wicke_molecular_2017} considered a 2D environment with uniform flow to study the impact of bounded drift-diffusion in more detail.
On the other hand, 1D diffusion with drift has been studied in \cite{srinivas_molecular_2012,kim_symbol_2014}.
However, it is not clear when such a simplified model is applicable in typical molecular communication application scenarios since flow in blood vessels or in microfluidic channels, i.e., in ducts, especially at the microscale, is far from uniform \cite{probstein_physicochemical_2005,bruus_theoretical_2007}.
Hence, in general, a reduction of the 3D reality to a 1D model is not justified.
In particular, the marginal axial and cross-sectional particle distributions are inherently coupled, which makes a mathematical analysis of the channel characteristics difficult.
This coupling has been considered in a heuristic parametric model in \cite{Kuscu_Modeling_2018} and simulated for blood vessels in~\cite{felicetti_simulation_2013}.

The notion of dispersion as the interaction of diffusion and non-uniform laminar flow was principally investigated in \cite{taylor_dispersion_1953,aris_dispersion_1956} and is now known as \emph{Taylor dispersion} \cite{probstein_physicochemical_2005}.
Via an \emph{effective diffusion coefficient}, the particle distribution can be derived in the regime of large source-observer distances where the interaction of cross-sectional diffusion and non-uniform flow yields a uniform particle distribution in the cross-section and a Gaussian spread along the axis.
For molecular communication, the authors in \cite{he_channel_2016,chahibi_pharmacokinetic_2015,bicen_system-theoretic_2013,sun_channel_2017,Bicen_Shannon_2018} adopted this model to keep their analysis analytically tractable but the conditions under which such simplifications are justified are not considered in detail in these works.
In particular, for short distances on the order of the duct radius, typically the impact of flow dominates as there is not enough time for diffusion to affect the overall particle distribution.
Recently, for molecular communication, this behavior which is in stark contrast to a diffusion regime, was also observed experimentally \cite{Unterweger_Experimental_2018}.
Thereby, the injection process can completely determine the entire channel characteristics.
There are several approaches for modeling the injection depending on the considered setup.
One general first-order model of the injection process is to assume a uniform initial particle distribution \cite{Levenspiel_Chemical_1999}.

The focus of this paper is twofold.
First, unlike previous works, we introduce the notion of dispersion in a systematic manner for molecular communication.
Second, we analyze and highlight the major effects of the advection-diffusion particle transport on molecular communication systems for two different regimes, namely the dispersion regime and the flow-dominated regime, the latter of which has not been considered in the molecular communication literature but is prevailing for example in blood vessels~\cite{probstein_physicochemical_2005}.
For this new regime, we derive the channel impulse response and the symbol error rate (SER) in on-off keying (OOK).

The two key results of this paper are as follows:
\begin{enumerate}
    \item
        There is a regime where one-dimensional diffusion with constant drift can accurately capture the channel characteristics by means of an effective diffusion coefficient and the cross-sectional mean velocity.
        In this regime, the initial spatial release pattern at the transmitter does not affect the particle distribution at the receiver.
    \item
        Non-uniform flow as encountered in ducts can cause significant intersymbol interference (ISI), especially in a flow-dominated regime.
        Diffusion tends to decrease ISI by enabling slowly-moving particles to move away from the boundary of the duct.
\end{enumerate}

The remainder of this paper is structured as follows.
In Section~\ref{sec:system_model}, we introduce the system model and present some preliminaries.
Section~\ref{sec:system_analysis} analyzes the duct channel and the different flow regimes.
Numerical results are presented in Section~\ref{sec:numerical}.
Finally, in Section~\ref{sec:conclusion}, we draw some conclusions.

\section{System Model and Preliminaries}
\label{sec:system_model}
\subsection{System Model}
\begin{figure}[!t]
    \centering
    \subfloat[]{
        \includegraphics[scale=0.8]{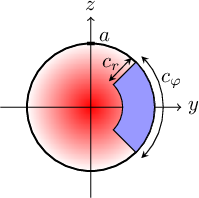}
    }
    \hfil
    \subfloat[]{
        \includegraphics[scale=0.8]{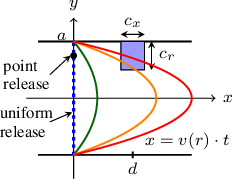}
        \label{fig:system_model_axial}
    }
    \caption{
        \label{fig:system_model}
        System model geometry (a) in the cross-section, and (b) along the axis.
        The red shading in (a) reflects the flow velocity which is maximum in the center and vanishes at the boundary.
        The corresponding parabolic shape, $x=v(r)\cdot t$, on which released particles reside when not diffusing after a uniform release, is sketched in (b) for three different time instances.
        Point and uniform transmitter release are shown as black dot and as a blue line, respectively.
        The receiver region is shaded in blue.
        \vspace*{-6mm}
    }
\end{figure}

We consider a straight impermeable cylindrical duct of infinite axial extent and radius $a$ which can be described by cylindrical coordinates $(x,r,\varphi)$, where $x\in(-\infty,\infty)$ is the axial position, $r\in[0,a]$ is the radial distance, and $\varphi\in(-\pi,\pi]$ is the azimuth angle.
The duct is filled with a fluid of viscosity $\eta$ that is subject to steady laminar flow in $x$-direction where the flow velocity $v(r)$ is a function of $r$ only and is given by a parabolic function; see Fig.~\ref{fig:system_model}.

At the transmitter (TX), we assume OOK modulation with a symbol interval length $T$, i.e., the data rate is $1/T$.
In particular, the TX releases $\Ntx$ and $0$ particles instantaneously at time $kT$ for transmitting a binary $1$ and $0$ in the $k$-th symbol interval, respectively.
Thereby, we consider releases either 1) uniformly and \emph{randomly} distributed over the cross section at $x=0$, or 2) from the point $(0,r_0,\varphi_0)$.
Moreover, we assume a transparent receiver (RX), which does not impact the particle transport.
Thereby, the RX is assumed to be able to detect and count the particles within its sensing volume specified by the points $(x,r,\varphi)$ satisfying $|x-d|\leq c_x/2, a-c_r\leq r\leq a, |\varphi|\leq c_\varphi/2$, i.e., the receiver is mounted on the duct wall with axial TX-RX distance $d$, radial extent $c_r$, and spanning an angle of $c_\varphi$, see Fig.~\ref{fig:system_model}.
Detection of the OOK symbols $a[k]\in\{0,1\}$ is performed based on the number of observed particles $\Nob(t)$ by applying a decision threshold $\xi\in\{0,1,\dots,\Ntx\}$:
\begin{equation}
    \label{eq:ahat}
    \ahat[k] =
    \begin{cases}
        0, & \Nob(t_0+kT) < \xi \\
        1, & \Nob(t_0+kT) \geq \xi,
    \end{cases}
\end{equation}
where $t_0$ is a detection delay and $\ahat[k]$ is the detected OOK symbol in the $k$-th symbol interval.

The released particles are transported by the fluid flow and Brownian motion.
As is usually done \cite{bruus_theoretical_2007}, we assume that particles do not interact with each other and do not influence the flow field.
Because of their small size, other forces such as gravity acting on the particles are negligible.

\subsection{Preliminaries}
\label{sec:preliminaries}
In molecular communication, information is conveyed by mass transfer.
Mass transfer in fluids is mediated by flow and Brownian motion which is referred to as \emph{advection} and \emph{diffusion}, respectively.
Thereby, mass transfer can be characterized by a time-varying spatial probability density function (PDF) $p(\bm{r};t)$ which can be interpreted as a normalized concentration where $\dV\cdot p(\bm{r};t)$ gives the average fraction of particles within the differential volume $\dV$ at $\bm{r}$ at time $t$.
The PDF $p(\bm{r};t)$ can be found as the solution to the following partial differential equation (PDE), which we will also refer to as the advection-diffusion equation \cite[Eq.~(5.22)]{bruus_theoretical_2007}
\begin{equation}
    \label{eq:advec-diffus}
    \partial_t p = D\nabla^2 p - \nabla \cdot p\bm{v}(\bm{r}),
\end{equation}
where $\partial_t p=\frac{\partial}{\partial t} p$ denotes the partial derivative of $p$ with respect to $t$ and $\nabla$ is the Nabla operator.
Moreover, $D$ is the diffusion coefficient and $\bm{v}(\bm{r})$ is the velocity vector at point $\bm{r}$.

To solve \eqref{eq:advec-diffus}, we need to know the velocity field $\bm{v}(\bm{r})$.
In general, the velocity field can be obtained by solving the Navier-Stokes equation, which provides a fundamental description of flow by relating the velocity field to the local pressure \cite[Ch.~2]{bruus_theoretical_2007}.
Applied to rigid and straight channels with no-slip boundary conditions, i.e., where the velocity at the boundary is zero, and subject to pressure-driven flow in the steady-state, the velocity profile is referred to as \emph{Poiseuille flow} \cite[Ch.~3]{bruus_theoretical_2007}.
Thereby, assuming a \emph{Newtonian fluid}, i.e., a fluid which can be described by the viscosity $\eta$, we can apply \cite[Eq.~(3.32)]{bruus_theoretical_2007}
\begin{equation}
    \label{eq:Hagen-Poiseuille}
    v(r) = 2\vmean \bigg(1-\frac{r^2}{a^2}\bigg),
\end{equation}
where $\vmean$ is the mean velocity in the channel and is a function of the applied pressure gradient $\pgrad$.
In particular, $\vmean$ can be obtained as \cite[Eq.~(3.34)]{bruus_theoretical_2007} $\vmean = |\pgrad| a^2/(8\eta)$.

We note that the maximum velocity $\vmax=2\vmean$ occurs at the center and can be found using $r=0$ in \eqref{eq:Hagen-Poiseuille}.

An important parameter for fluid flow is the P\'eclet number, which characterizes the relative importance of diffusion and advection.
This dimensionless number is defined as \cite[Eq.~(4.6.8)]{probstein_physicochemical_2005}
\begin{equation}
    \label{eq:Peclet}
    \Pe = \frac{\vmean\, a}{D}.
\end{equation}

Intuitively, $\Pe>0$ increases and decreases when $\vmean$ and $D$ increase as the importance of particle transport by flow and diffusion becomes more relevant, respectively.

\vspace*{-2mm}
\section{Analysis of the Duct Channel}
\label{sec:system_analysis}
The advection-diffusion equation \eqref{eq:advec-diffus} for the environment in Fig.~\ref{fig:system_model} simplifies to the following PDE:
\begin{IEEEeqnarray}{rCl}
    \label{eq:sys_pde}
    \partial_t p &=& D\nabla^2 p - v(r)\partial_x p,
\end{IEEEeqnarray}
for $t>0$ because the velocity field is independent of the axial position.
At the boundary $r=a$, $\partial_r p = 0$ has to hold and $p(x,r,\varphi;0)$ is initially given by $\delta(x)/(\pi a^2)$ and $\delta(x)\delta(r-r_0)\delta(\varphi-\varphi_0)/r$ for uniform and point release, respectively.
Eq.~\eqref{eq:sys_pde} is still difficult to solve in general because of the nonlinear velocity \eqref{eq:Hagen-Poiseuille} and the inherent coupling of $p$ in the $x$- and $r$-directions.
Nonetheless, in certain parameter regimes, \eqref{eq:sys_pde} can be solved in closed-form.

For these regimes, we seek the time-dependent observation probability
\begin{equation}
    \label{eq:pob_int}
    \Pob(t) = \int_{\Vrx} p(x,r,\varphi;t)\,\mathrm{d}\Vrx,
\end{equation}
where $\Vrx$ is the RX volume.
We will also refer to $\Pob(t)$ as the \emph{impulse response} of the molecular communication channel.
The impulse response is a fundamental characteristic of the molecular communication channel as it determines the mean $\Nobmean(t)$ of the received signal $\Nob(t)$ \cite{noel_improving_2014}.

In the following, we will investigate two special regimes for which \eqref{eq:pob_int} can be solved analytically \cite[Chapter~4.6]{probstein_physicochemical_2005}:
\begin{IEEEeqnarray}{stl}
    dispersion regime & for & \vmean a/D\ll 4d/a\label{eq:Peclet_limit}\\
    flow-dominated regime & for & \vmean a/D\gg 4d/a. \label{eq:Peclet_flow}
\end{IEEEeqnarray}

Intuitively, Eq.~\eqref{eq:Peclet_limit} (Eq.~\eqref{eq:Peclet_flow}) states that the time $d/\vmean$ required for particles to be transported by flow with mean velocity $\vmean$ over distance $d$ in $x$-direction is much larger (smaller) than $a^2/(4D)$, which is characteristic for free diffusion over distance $a$, i.e., within the bounded domain in the $y$-$z$-plane, diffusion has had (has not had) enough time to interact with the non-uniform flow-profile.
Naturally, \eqref{eq:Peclet_limit} and \eqref{eq:Peclet_flow} include the special cases of pure diffusion where flow is not present (i.e., $\vmean=0$) and flow-only transport when there is no diffusion (i.e., $D=0$), respectively.
\subsection{Dispersion Regime}
Dispersion is the result of the interaction of cross-sectional diffusion and non-uniform advection due to the flow profile.
This interaction can lead to a particle distribution that is uniform in each cross-section, i.e., the spatial PDF can be written as $p(x,r,\varphi;t)=p(x;t)/(\pi a^2)$.
In this regime, the particle distribution does depend only on the initial $x$-position, i.e., there is no difference between point and uniform release at the TX.
We can rearrange \eqref{eq:Peclet_limit} as
\begin{equation}
    \label{eq:D_limit}
    D \gg \frac{a^2\cdot\vmean}{4d},
\end{equation}
which characterizes the diffusion coefficients required for dispersion to occur.
If \eqref{eq:Peclet_limit} is satisfied, \eqref{eq:sys_pde} can be written as the following 1D advection-diffusion equation~\cite[Eq.~(4.6.30)]{probstein_physicochemical_2005}
\begin{equation}
    \label{eq:res_pde}
    \partial_t p = \Deff\, \partial_x^2 p - \vmean \,\partial_x p,
\end{equation}
with \emph{effective diffusion coefficient} $\Deff$ and mean velocity $\vmean$.
For an instantaneous uniform release at $x=0$, the solution to \eqref{eq:res_pde} is given by
\begin{equation}
    \label{eq:pde_sol}
    p(x,r,\varphi;t) = \frac{1}{\pi a^2}\times\frac{1}{\sqrt{4\pi \Deff\, t}} \,{\exp}{\left(-\frac{(x-\vmean t)^2}{4\Deff\, t}\right)},
\end{equation}
where $1/(\pi a^2)$ represents the uniform cross-sectional distribution in the $y$-$z$-plane.

Following \cite[Eq.~(26)]{aris_dispersion_1956}, the \emph{Taylor-Aris} effective diffusion coefficient $\Deff$ is obtained as \cite[Eq.~(4.6.35)]{probstein_physicochemical_2005}
\begin{equation}
    \label{eq:Deff}
    \Deff = D\left(1 + \frac{1}{48}\left(\frac{\vmean\, a}{D}\right)^2\right).
\end{equation}

We note that in general $\Deff>D$ and moreover $\Deff\gg D$ when $D$ is decreased to very small values, which by \eqref{eq:Peclet} increases $\Pe$.
However, by \eqref{eq:Peclet_limit}, decreasing $D$ comes at the expense of a larger required distance for dispersion to take place, cf. \eqref{eq:D_limit}.

Employing \eqref{eq:pob_int}, the observation probability obtained by integrating \eqref{eq:pde_sol} over the receiver volume is
\begin{IEEEeqnarray}{rCl}
    \label{eq:pob_diffus_sol}
    \Pobd(t) &= &\frac{\Arx}{a^2} \times \bigg[{\qfunc}{\left(\frac{d-c_x/2-\vmean\,t}{2\Deff\, t}\right)}\nonumber \\
             & &-\, {\qfunc}{\left(\frac{d+c_x/2-\vmean\,t}{2\Deff \,t}\right)}\bigg],
\end{IEEEeqnarray}
where $\qfunc(\cdot)$ is the Gaussian Q-function and $\Arx=c_\varphi/(2\pi) \cdot (2ac_r - c_r^2)$.

It is of interest to derive the time at which $\Pobd(t)$ attains its maximum as this may serve as design guideline for the symbol interval length.
As maximizing $\Pobd(t)$ with respect to $t$ is cumbersome, we resort to maximizing \eqref{eq:pde_sol} for $x=d$, which yields
\begin{equation}
    \label{eq:peak_time}
    \tmax = \frac{\Deff}{\vmean^2}\left(-1+\sqrt{1+\frac{\vmean^2}{\Deff^2} d^2}\,\right).
\end{equation}

In this approximation, the peak height follows as $\pmax = \Pobd(\tmax)$.
We note that because of diffusion $\tmax<d/\vmean$, where $d/\vmean$ is the time when particles moving with the mean velocity will reach the RX.

\subsection{Flow-dominated Regime}
In this subsection, we determine the observation probability $\Pobf(t)=\Pob(t)$ in regime \eqref{eq:Peclet_flow}.

\subsubsection{Uniform Release}
First, we assume a uniform release at $x=0$.
In this case, all particles will lie on the surface of a paraboloid that extends along the axis over time and exhibits rotational symmetry.
Thereby, the marginal distribution in the cross-section, specified by the $r$- and $\varphi$-coordinates, remains uniform because the flow is in the $x$-direction.
In this case, the spatial distribution can be written as
\begin{equation}
    \label{eq:pde_sol_advec}
    p(x,r,\varphi;t) = \frac{1}{\pi a^2} \delta(x - v(r)\cdot t),
\end{equation}
with $v(r)$ given in \eqref{eq:Hagen-Poiseuille}.

By integrating \eqref{eq:pde_sol_advec} over the receiver volume, we can derive the impulse response in the flow-dominated regime following a uniform release:
\begin{equation}
    \label{eq:pob_advec_sol}
    \Pobf(t) =
    \begin{dcases}
        0, & t\leq t_1 \\
        \left[\frac{1}{a^2}\Arx-\frac{c_\varphi}{2\pi}\frac{d-c_x/2}{2\vmean t}\right], & t_1<t<t_2 \\
        \frac{c_\varphi}{2\pi} \cdot \frac{c_x}{2\vmean t}, & t\geq t_2.
    \end{dcases}
\end{equation}
where
\begin{IEEEeqnarray}{rCl}
    \label{eq:t2}
    t_{1,2} &=& \frac{d\mp c_x/2}{2\vmean(1-(1-c_r/a)^2)}.
\end{IEEEeqnarray}

$\Pobf(t)$ is maximized for $t=t_2$.
Also, at time $t=t_2/\alpha$ the fraction $\alpha\in(0,1]$ of $\Pobf(t_2)$ can be observed.
The tail of the impulse response decays only polynomially with time, which may give rise to significant ISI in molecular communication systems.

\subsubsection{Point Release}
For a point release with $r_0\in[a-c_r,a]$ and $\varphi_0\in[-c_\varphi/2,c_\varphi/2]$, i.e., when the TX coordinates are within the $r$- and $\varphi$-coordinates of the RX, we observe all particles with certainty if $d-c_x/2\leq v(r_0) t\leq d+c_x/2$, i.e., the impulse response is given by
\begin{equation}
    \label{eq:pob_advec_point}
    \Pobfp(t) = {\rect}{\left(\frac{v(r_0)\, t - d}{c_x}\right)},
\end{equation}
where $\rect(x) = 1$ if $-1/2\leq x\leq 1/2$ and zero otherwise.
When the release point is \emph{not} within the $r$- and $\varphi$-coordinates of the RX then the impulse response is zero for all times.

We note that \eqref{eq:pob_advec_sol} and \eqref{eq:pob_advec_point} still give the observation \emph{probability} $\in[0,1]$ even though the flow is deterministic, i.e., with probability $\Pob(t)$ and $1-\Pob(t)$ each of the $\Ntx$ particles can be independently and cannot be observed within the RX volume at time $t$, respectively.
The reason for this is that for both uniform and point release the initial particle position can be understood as independently and uniformly random distributed within the available TX area (the whole cross section or one point).

\subsection{Performance Metrics}
Using $\Pob(t)$, for the OOK symbol sequence $a[k], k\in\{0,1,\dots,K-1\}$, of length $K$, we can easily determine the expected number of observed particles at the RX as
\begin{equation}
    \Nobmean(t) = \Ntx \sum_{k=0}^{K-1} a[k] \Pob(t-kT) + \Nn,
\end{equation}
where $\Nn$ is the average number of received external noise molecules.
We note that, in general, $\Nob(t)$ is a binomial random variable even for $D\to0$ because of the random initial distribution for each particle release.

Consequently, the \emph{average} SER can be defined as
\begin{equation}
    \label{eq:ser}
    P_\mathrm{e} = \frac{1}{2^K}\sum_{a\in\mathbb{A}(K)} \left[\frac{1}{K}\sum_{i=0}^{K-1} {\Pr}{\left(\hat{a}[i]\neq a[i];\; a[j\leq i]\right)} \right],
\end{equation}
where $\mathbb{A}(K)$ is the set of all $2^K$ possible binary sequences of length $K$ and the expression ${\Pr}{\left(\hat{a}[i]\neq a[i];\; a[j\leq i]\right)}$ represents the probability of detecting $a[i]$ incorrectly given the transmitted sequence $a[j]$ for $0\leq j\leq i$.

For computational convenience, the binomial distribution for $\Nob(t)$ can be approximated by a Poisson distribution with mean $\Nobmean(t)$ when $\Ntx$ is sufficiently large and $\Pob(t_0)$ is sufficiently small \cite{noel_improving_2014}.
Using this Poisson approximation, ${\Pr}{\left(\hat{a}[i]\neq a[i];\; a[j\leq i]\right)}$ is readily obtained as \cite[Eq.~(31)]{noel_improving_2014}.
The accuracy of this approximation for the considered system parameters is validated in Section~\ref{sec:numerical}.

\section{Numerical Evaluation}
\label{sec:numerical}
By using particle-based simulation, we validate our derived analytical expressions and explore those regimes for which mathematical analysis is not readily accomplished.
Thereby, unless explicitly stated otherwise, we employ the following physical parameter values.
As diffusion coefficient we choose $D=\SI{e-10}{\meter^2\per\second}$ which is a reasonable estimate for small proteins \cite{probstein_physicochemical_2005}.
Two values for the duct radius, $a=\SI{10}{\micro\meter}$ and $a=\SI{200}{\micro\meter}$, are considered, which is reasonable for small capillaries \cite{probstein_physicochemical_2005} and microfluidic ducts \cite{bruus_theoretical_2007}, respectively.
Furthermore, two TX-RX distances are considered with values $d=\SI{200}{\micro\meter}$ and $d=\SI{800}{\micro\meter}$.
Moreover, we choose the receiver dimensions as $c_x=a/2$, $c_r=a/2$, $c_\varphi=\pi/2$, i.e., the receiver size scales with the duct radius.
The microscopic simulation time step is set to $\Delta t=\SI{e-3}{\second}$.
The fluid flow mean velocity is assumed to be $\vmean = \SI{1}{\milli\meter\per\second}$, which is reasonable for small capillaries~\cite{probstein_physicochemical_2005}.
For $a=\SI{10}{\micro\meter}$ and $D=\SI{e-10}{\meter^2\per\second}$, we obtain $\Deff=\SI{2.1e-8}{\meter^2\per\second}$, which is a value otherwise unattainable for the diffusion coefficient of small proteins \cite{probstein_physicochemical_2005}.

\begin{figure}[!t]
    \centering
    \includegraphics[scale=0.8]{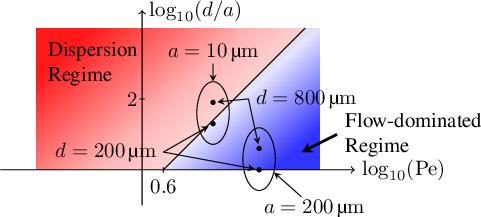}
    \figspaceupper
    \caption{
        \label{fig:regions}
        Sketch of regions of different transport regimes.
        Adapted from \cite{probstein_physicochemical_2005}.
        All four simulation scenarios are shown as black dots.
        For $a=\SI{10}{\micro\meter}$, we have $\Pe=100$ and $d/a=20,80$.
        For $a=\SI{200}{\micro\meter}$, we have $\Pe=2000$ and $d/a=1,4$.
        \vspace*{-1em}
    }
\end{figure}

We show in Fig.~\ref{fig:regions} (adapted from \cite{probstein_physicochemical_2005}) the considered parameter values in terms of the P\'eclet number and the ratio of the TX-RX distance to the duct radius.
In particular, we have shaded the two regimes for which the obtained analytical results in Section~\ref{sec:system_analysis} are expected to be applicable.
By \eqref{eq:Peclet_limit} and \eqref{eq:Peclet_flow}, $\Pe=4d/a$ separates these two regimes and is shown as a black line.
The derived analytical results are valid for parameter values well within the dispersion or the flow-dominated regions.
However, the analytical results cannot be expected to be accurate close the boundary shown by the black line.
For the two duct radii $a=\SI{10}{\micro\meter},\SI{200}{\micro\meter}$ and the two TX-RX distances $d=\SI{200}{\micro\meter},\SI{800}{\micro\meter}$, we show the resulting four combinations of $d/a$ and $\Pe$ as black dots in Fig.~\ref{fig:regions}.
We see that the two scenarios for $a=\SI{10}{\micro\meter}$ lie close to the boundary of both regimes.
These parameter values have been chosen such that simulations can reveal the deviations from either regime.
On the other hand, the scenarios for $a=\SI{200}{\micro\meter}$ lie well within the flow-dominated regime and we expect no deviations from the developed theory.
We note that changing the duct radius $a$ affects both $d/a$ and $\Pe$ whereas a change in $d$ influences only $d/a$.
Considering the parameter values chosen in this paper, from Fig.~\ref{fig:regions}, we can conclude that the dispersion regime is most applicable for small microscale ducts.
On the other hand, we also see that there is a large set of parameters for which the flow-dominated regime is more appropriate, especially for medium to large ducts.

\begin{figure}[!t]
    \centering
    \includegraphics[scale=0.8]{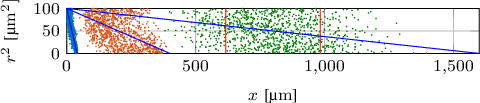}
    \figspaceupper
    \caption{
        \label{fig:snapshot}
        Snapshot of particle positions for $a=\SI{10}{\micro\meter}$ and at $t=0.02,0.2,\SI{0.8}{\second}$ after uniform release at $x=0$ and $t=0$ shown in different colors and starting from left to right, respectively.
        In total, $\Ntx=\num{e3}$ are released.
        \vspace*{-4mm}
    }
\end{figure}

To gain a basic understanding of the particle evolution towards dispersion, in Fig.~\ref{fig:snapshot}, we show three snapshots of the particle positions corresponding to three different time instances and distinguished by different colors following a uniform release at $t=0$.
In particular, we plot $r^2$ over $x$ motivated by the fact that the marginal distribution in $r^2$ of a uniform distribution within a circular disk is uniform.
As a side effect, for the flow-dominated regime, via \eqref{eq:pde_sol_advec} $r^2$ becomes a simple linear function of $x$ which for each considered $t$ is shown as a blue line.
For the largest time shown, $t=\SI{0.8}{\second}$, the red lines show the standard deviation positions $\vmean t \pm \sqrt{2\Deff\, t}$ from the mean when assuming the Gaussian distribution in \eqref{eq:pde_sol} due to dispersion.
For small times, e.g., $t=\SI{0.02}{\second}$, the particles follow the parabolic profile closely.
For slightly larger times, e.g., $t=\SI{0.2}{\second}$, the particles start to spread because of diffusion.
For large times, e.g., $t=\SI{0.8}{\second}$, particles become uniformly distributed along the $r^2$ dimension within the duct due to dispersion.
In summary, when considering the mean particle position $\vmean\cdot t$, at small times after the release the flow-dominated regime and at large times after the release the dispersion regime accurately model the actual behavior.

\begin{figure}[!t]
    \centering
    \subfloat[]{
        \includegraphics{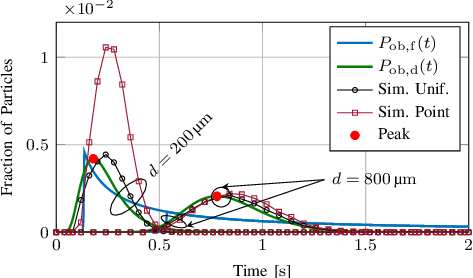}
        \label{fig:ir_small_diameter}
    }

    \figspacelower
    \subfloat[]{
        \includegraphics{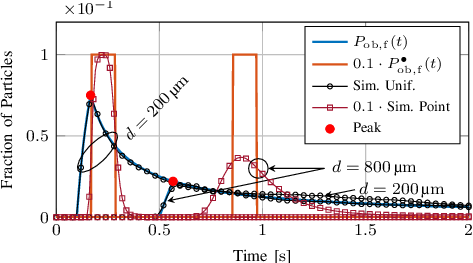}
        \label{fig:ir_large_diameter}
    }
    \caption{
        Impulse responses for (a) $a=\SI{10}{\micro\meter}$, and (b) $a=\SI{200}{\micro\meter}$.
        Simulation results are shown for both uniform and point release.
        For (b), additionally the simulated and analytical impulse responses due to a point release are scaled by the constant factor \num{0.1} for a better visualization.
        Simulation results are averaged using $\Ntx=\num{e6}$.
        \vspace*{-5mm}
    }
\end{figure}

In Figs.~\ref{fig:ir_small_diameter} and \ref{fig:ir_large_diameter}, we show the impulse response for $a=\SI{10}{\micro\meter}$ and $a=\SI{200}{\micro\meter}$, respectively.
In each case, we consider both $d=\SI{200}{\micro\meter}$ and $d=\SI{800}{\micro\meter}$ as well as uniform and point release.
For the point release, the position $(0,0.75 a,0)$ was chosen such that particles arrive at the receiver when not diffusing.
We simulate the impulse responses and investigate which of the developed analytical models provides the best fit in each case.

In Fig.~\ref{fig:ir_small_diameter}, for comparison, both $\Pobd(t)$ (applicable for both point and uniform release) in \eqref{eq:pob_diffus_sol} and $\Pobf(t)$ (applicable only for uniform release) in \eqref{eq:pob_advec_sol} are shown.
Thereby, for $\Pobd(t)$, the peaks via \eqref{eq:peak_time} are also highlighted with large dots.
When $d=\SI{200}{\micro\meter}$, the simulated impulse response following a point release is significantly larger than that for a simulated uniform release.
In this case, both simulated impulse responses neither match $\Pobd(t)$ nor $\Pobf(t)$.
However, especially considering the long-time behavior, (e.g., for $t>\SI{0.5}{\second}$) the simulated data points tend to be better described by $\Pobd(t)$ than by $\Pobf(t)$.
For $d=\SI{800}{\micro\meter}$, the deviations of the simulated impulse responses for point and uniform release from $\Pobd(t)$ are much smaller and the dispersion regime provides a much better fit despite the fact that \eqref{eq:Peclet_limit} is not strictly satisfied, cf. Fig.~\ref{fig:regions}.
This is consistent with the green particle cloud in Fig.~\ref{fig:snapshot} which appears uniform in $r^2$.
Comparing $\Pobf(t)$ and $\Pobd(t)$, we see that the peak of $\Pobf(t)$ is larger and smaller than that of $\Pobd(t)$ when $d$ is small and large, respectively.
For larger times, e.g., for $t>\SI{0.75}{\second}$, $\Pobf(t)$ for $d=\SI{200}{\micro\meter}$ and $\Pobf(t)$ for $d=\SI{800}{\micro\meter}$ coincide as expected from \eqref{eq:pob_advec_sol} which is independent of $d$ for $t>t_2$.
In conclusion, parameter values close to the boundary in Fig.~\ref{fig:regions} can still be applicable for the dispersion model.

In Fig.~\ref{fig:ir_large_diameter}, $\Pobf(t)$ in \eqref{eq:pob_advec_sol} and $\Pobfp(t)$ in \eqref{eq:pob_advec_point} are shown.
For the former, the peak times $t_2$ are highlighted.
For both uniform and point release, simulation results are also shown.
By Fig.~\ref{fig:regions}, the dispersion approximation is not applicable in this scenario and for clarity is not shown.
Considering the point release, the simulated curve for $d=\SI{200}{\micro\meter}$ matches the rectangular shape of $\Pobfp(t)$ in \eqref{eq:pob_advec_point} reasonably well.
On the other hand, for $d=\SI{800}{\micro\meter}$, the simulated impulse response significantly deviates from the rectangular shape because diffusion has had enough time to result in a spread of the pulse.
As expected from Fig.~\ref{fig:regions}, we observe in general a good agreement between $\Pobf(t)$ in \eqref{eq:pob_advec_sol} and the simulation results in the case of a uniform release, i.e., the flow-dominated regime provides a reasonable description of the channel.
Nevertheless, for $d=\SI{200}{\micro\meter}$, at larger times (e.g., for $t>\SI{1}{\second}$), there is a small deviation because of residual particles close to the RX when most particles have already passed.
Comparing the impulse responses for point release and uniform release, we find that the tail of the impulse response strongly depends on the initial distribution.
There can be considerable ISI, especially when a fraction of the particles is released close to the duct wall.
However, as long as the RX can be reached by the particles, an initial release close to the center of the duct might reduce~ISI.

Comparing Figs.~\ref{fig:ir_small_diameter} and \ref{fig:ir_large_diameter}, we see that the channels behave quite differently when the duct radius is changed from $a=\SI{10}{\micro\meter}$ to $a=\SI{200}{\micro\meter}$.
We note that for both uniform and point release the peak values of the impulse responses in Fig.~\ref{fig:ir_small_diameter} are at least by an order of magnitude smaller than in Fig.~\ref{fig:ir_large_diameter}.
Moreover, for uniform and point release the simulated impulse responses in Fig.~\ref{fig:ir_small_diameter} decay faster and slower from their peak values than those in Fig.~\ref{fig:ir_large_diameter}, respectively.

\begin{figure}[!t]
    \centering
    \includegraphics[scale=0.8]{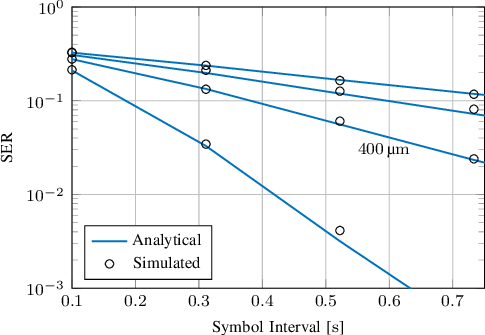}
    \vspace*{-3mm}
    \caption{
        Symbol error rate in \eqref{eq:ser} as a function of the symbol interval length for $d=200,400,600,\SI{800}{\micro\meter}$.
        Parameters are chosen as $\Ntx=\num{e3}$, $\Nn=4$, $D=\SI{e-12}{\meter^2\per\second}$, and $K=8$.
        Results are averaged over \num{e4} independent realizations.
        \vspace*{-1em}
    }
    \label{fig:ser}
\end{figure}

In Fig.~\ref{fig:ser}, we show the average SER in \eqref{eq:ser} as a function of the symbol duration $T$ employing the analytical impulse response in \eqref{eq:pob_advec_sol} for a uniform release in the flow-dominated regime.
The analytical results are validated by particle-based simulation which match reasonably well.
Deviations shown may be attributed to the averaging.
To accurately capture the flow-dominated regime, we choose $D=\SI{e-12}{\meter^2\per\second}$ and $a=\SI{200}{\micro\meter}$.
The curves are parameterized by the TX-RX distance $d$, which is varied from \SI{200}{\micro\meter} to \SI{800}{\micro\meter}.
As detection time offset, i.e., as delay, $t_0=t_2$ in \eqref{eq:t2} is chosen and an average of $\Nn=\num{4}$ noise molecules is assumed.
For each considered symbol interval and TX-RX distance, the optimal threshold $\xi$ minimizing the SER is found by full search and employed.
In general, the SER decreases for increasing $T$ at the expense of decreasing the data rate $1/T$.
However, for moderate to large distances, e.g., $d=\SI{800}{\micro\meter}$, the SER does not decrease significantly even if the symbol interval is relatively large, e.g., $T=\SI{0.75}{\second}$, because of severe ISI, cf. Fig.~\ref{fig:ir_large_diameter}.
The SER for larger distances could potentially be decreased by additional equalization or by adapting the injection mechanism as can drastically be seen by the rectangular-shape impulse response for a point release in Fig.~\ref{fig:ir_large_diameter} which can exhibit negligible ISI.

\section{Conclusion}
\label{sec:conclusion}
Transport by non-uniform fluid flow and diffusion can be categorized into different regimes, depending on the relative importance of the two transport phenomena.
Two extreme (but widely applicable) regimes are the dispersion and the flow-dominated regimes.
For a given duct radius, either regime can be applicable depending on the TX-RX distance and the P\'eclet number.
Dispersion generalizes the concept of diffusion which is crucial for signal propagation in molecular communication.
Thereby, a non-uniform flow can be accounted for by an effective diffusion coefficient.
This effective diffusion coefficient can be multiple orders of magnitude larger than the actual diffusion coefficient.
However, this description relies on a large TX-RX distance.
On the other hand, there are many practical scenarios at the microscale that fall within a flow-dominated regime where dispersion is insignificant.
In this regime, the initial release pattern drastically influences the ISI.

\section*{Acknowledgment}
This work was supported in part by the Friedrich-Alexander University of Erlangen-N\"urnberg (FAU) under the Emerging Fields Initiative (EFI) and the STAEDTLER Foundation.



\bibliographystyle{IEEEtran}
\bibliography{IEEEabrv,main.bbl}
%

\end{document}